\documentclass[preprint,showpacs,aps,prd]{revtex4-1}
\usepackage{epsfig}
\usepackage{psfrag}
\usepackage{graphicx}
\usepackage{titlesec}
\usepackage{graphics}
\usepackage{amsmath}
\usepackage{color}
\usepackage[title]{appendix}

\begin{document}


\title{On a Gambini-Pullin-inspired Maxwell electrodynamics with Lorentz-violating dimension-$5$ operators\footnote{This work is dedicated to the memory of our friend and collaborator Antonio Aurilia.}}

\author{Patricio Gaete} \email{patricio.gaete@usm.cl} 
\affiliation{Departamento de F\'{i}sica and Centro Cient\'{i}fico-Tecnol\'ogico de Valpara\'{i}so-CCTVal,
Universidad T\'{e}cnica Federico Santa Mar\'{i}a, Valpara\'{i}so, Chile}

\author{Jos\'e Abdalla Helay\"{e}l-Neto}\email{helayel@cbpf.br}
\affiliation{Centro Brasileiro de Pesquisas F\'{i}sicas (CBPF), Rio de Janeiro, RJ, Brasil}

\begin{abstract}
We explore the properties of a new Maxwell electrodynamics  coupled to a Lorentz-violating background through the presence of higher-derivative terms. Physical implications of this alternative effective theory modified by Lorentz-violating operators of mass dimension-$5$ are considered, such as modified dispersion relations which exhibit the vacuum birefringence phenomenon. Subsequently, we analyze the energy-momentum tensor in the case we reproduce the Gambini-Pullin electrodynamics. We explicitly show that for a background time-like four-vector, $n^{\mu}$, the equations of motion indeed reproduce those encountered by Gambini and Pullin in the framework of loop quantum gravity . Finally, extended expressions for Maxwell-type equations are written down in the case the external Lorentz-violating four-vector is space-like.
\end{abstract}


\maketitle

\section{Introduction}

As it is well known, the Standard Model (SM) describes the interactions among the fundamental constituents of the matter, that is, the quantum field theory of the smallest building blocks of nature interacting in a locally gauge-invariant way. Its success is based on predicting with an impressive accuracy every process in the accelerators to this day, in a wide range of energy. Despite this amazing success, the SM is far from perfect. The problems begin, as widely discussed, with the fact that the SM does not include gravity, an interaction for which we do not have to this date a quantum description in four space-time dimensions. In fact, this has led to assume that these two theories will merge at the Planck scale, ${m}_{P}\cong{10}^{19}{GeV}$, in order to provide a consistent framework to unify all fundamental interactions \cite{Kostelecky:2003fs}.

Let us also mention here that, at these energies, Lorentz- and CPT-symmetry violation effects are expected to show up. Actually, the search for departures from Lorentz invariance is an active research area. The interest in studying these issues is mainly due to that in some candidate theories of a quantum description of gravity, such as string theory \cite{Kostelecky:1988zi,Kostelecky:1989jp,Kostelecky:1989nt,Kostelecky:1990pe} and loop quantum gravity \cite{Gambini:1998it,Alfaro:2001rb}, there are scenarios where the breakdown of Lorentz symmetry takes place. We mention, in passing, that the main idea of these developments is that there would be atoms of geometry of which  fundamental building blocks of gravity arise (discreteness). However, independent from the quantum gravity theory, the focus of these studies lies on low-energies, where the effects of Lorentz-symmetry violation (LSV) can be unveiled. This discreteness would act as a dispersive medium for particles, such as photons, for example, propagating on this spacetime.

In this connection, it becomes of interest, to recall that low-energy effective field theories are a very suitable framework to probe effects of quantum gravity introduced as small (LSV) terms in the usual field theories. The Standard Model Extension (SME) is the key example, so that possible departures from exact Lorentz invariance can be implemented. In fact, this has provided a theoretical background from different points of view. For example, in connection with photon physics \cite{phyphotons}, also in effects of radiative corrections \cite{radiat}, systems of fermions \cite{fermi}, neutrino physics \cite{neutrinos}, topological deffects \cite{defectop}, cosmic rays \cite{cosmic}, and in magnetic monopoles studies \cite{Turcati:2018bxj}. As well as, Lorentz-violating theories with higher-dimensional operators \cite{Kostelecky:2009zp,Myers:2003fd,Casana:2018rhg}.
Another possible scenario is a newly proposed theory of Lorentz-symmetry violation from first principles \cite{ Lingli:2011fd}. In addition, in previous studies \cite{Belich:2013rma,Belich:2015qxa}, we have considered aspects of LSV physics in a supersymmetric scenario. On the one hand, we have studied the minimal supersymmetric extension of the Carroll-Field-Jackiw model for Electrodynamics with a topological Chern-Simons-like LSV term. In this case, we have shown that supersymmetry is naturally broken whenever a single scalar supermultiplet triggers Lorentz violation symmetry. On the other hand, we have discussed that if the scale of breaking of Lorentz covariance is above the supersymmetry breaking scale, we find that once Lorentz symmetry is violated, supersymmetry also takes place. In this case the photon and photino dispersion relations are split from one another with different profiles in terms of the supersymmetric background induced in the process of LSV.

As we have already expressed, a particularly interesting example that spacetime would be discrete, within loop quantum gravity, is the approximation developed in \cite{Gambini:1998it,Alfaro:2001rb} for the Einstein-Maxwell theory. In fact, these authors devised a method for obtaining effective Maxwell equations from the expectation values of the quantum Hamiltonian. As a consequence, they obtain modified dispersion relations which display the breaking of Lorentz symmetry. Interestingly, a similar result has been obtained in the context of an effective theory with Lorentz violating dimension-$5$ operators for the photon sector \cite{Myers:2003fd}. Inspired by the above observations, in this work we would like to recover the results obtained in \cite{Gambini:1998it,Alfaro:2001rb} within the familiar language of standard (effective) quantum field theory in four dimensions. This piece of information can be useful in view of establishing new connections among different models describing the same physical phenomena. 

In Sect.2, we propose and describe this modified Maxwell electrodynamics with Lorentz-violating dimension-$5$ operators and study aspects of birefringence and the energy-momentum tensor.
In Sect.3, we study Maxwell-type equations in the case of a space-like four-vector. Finally, our Concluding Remarks are cast in Sect.4.

\section{The model under consideration}

As already mentioned in the introduction, in this section we will propose and study the properties of this modified Maxwell electrodynamics in the language of usual field theory. With this purpose, let us consider first a new electromagnetic field strength tensor. This would not only provide the setup theoretical for our subsequent work, but also fix the notation. 

\subsection{Electromagnetic field-strength tensor}

We start off our analysis by considering a new four-dimensional spacetime electromagnetic field-strength tensor here extended as follows below 
\begin{equation}
{\cal F}_{\mathit{\mu}\mathit{\nu}} = {F}_{\mathit{\mu}\mathit{\nu}} + \frac{\mathit{\alpha}}{2}\mathit{{{l}_{P} }}{n}_{\mathit{\kappa}}{\varepsilon}^{\mathit{\kappa}}{}_{\mathit{\mu}\mathit{\alpha}\mathit{\beta}}{\partial}^{\mathit{\alpha}}{F}_{\mathit{\nu}}{}^{\mathit{\beta}} - \frac{\mathit{\alpha}}{2}\mathit{{l}_{P}}{n}_{\mathit{\kappa}}{\varepsilon}^{\mathit{\kappa}}{}_{\mathit{\nu}\mathit{\alpha}\mathit{\beta}}{\partial}^{\mathit{\alpha}}{F}_{\mathit{\mu}}{}^{\mathit{\beta}} + \mathit{\beta}\mathit{{l}_{P}}{n}_{\mathit{\kappa}}{\varepsilon}^{\mathit{\alpha}}{}_{\mathit{\mu}\mathit{\nu}\mathit{\beta}}{\partial}_{\mathit{\alpha}}{F}^{\mathit{\kappa}\mathit{\beta}}, \label{G_H05}
\end{equation}
where ${l}_{P}$ is Planck's length, whereas $\alpha$ and $\beta$ are dimensionless numerical coefficients. Here, ${F}_{\mathit{\mu}\mathit{\nu}} = {\partial}_{\mathit{\mu}}{A}_{\mathit{\nu}} - {\partial}_{\mathit{\nu}}{A}_{\mathit{\mu}}$, stands for the usual electromagnetic field-strength tensor, and ${n}^{\mathit{\mu}}$ is a dimensionless four-vector.

From this equation, it follows that the dual electromagnetic field-strength tensor takes the form
\begin{eqnarray}
{ \tilde{{\cal F}}}^{\mathit{\mu}\mathit{\nu}} &=& {\tilde{F}}^{\mathit{\mu}\mathit{\nu}} - \mathit{{l}_{P}}\left({{-}\frac{\mathit{\alpha}}{2} + \mathit{\beta}}\right){n}_{\mathit{\kappa}}{\partial}^{\mathit{\mu}}{F}^{\mathit{\kappa}\mathit{\nu}} + \mathit{{l}_{P}}\left({{-}\frac{\mathit{\alpha}}{2} + \mathit{\beta}}\right){n}_{\mathit{\kappa}}{\partial}^{\mathit{\nu}}{F}^{\mathit{\kappa}\mathit{\mu}} \nonumber\\
&+&\mathit{{l}_{P}}\frac{\mathit{\alpha}}{2}{n}^{\mathit{\nu}}{\partial}_{\mathit{\kappa}}{F}^{\mathit{\kappa}\mathit{\mu}} - \mathit{{l}_{P}}\frac{\mathit{\alpha}}{2}{n}^{\mathit{\mu}}{\partial}_{\mathit{\kappa}}{F}^{\mathit{\kappa}\mathit{\nu}}, \label{G_H10}
\end{eqnarray}
where ${\tilde{F}}^{\mathit{\mu}\mathit{\nu}} = \frac{1}{2}{\varepsilon}^{\mathit{\mu}\mathit{\nu}\mathit{\rho}\mathit{\lambda}}{F}_{\mathit{\rho}\mathit{\lambda}}$ is the usual dual electromagnetic field-strength tensor.

For the sake of completeness, these equations can be alternatively written in terms of electric  (${\vec E}$) and magnetic (${\vec B}$) fields, as
\begin{equation}
{{\cal F}}_{0i} = {E}_{i}{-}\left({{-}\frac{\mathit{\alpha}}{2} + \mathit{\beta}}\right){\mathit{{l}_{P}}\varepsilon}_{ijk}{\partial}_{j}{E}_{k},  \label{G_H15}
\end{equation}

\begin{equation}
{{\cal F}}_{ij} = - {\varepsilon}_{ijk}\left({{B}_{k} - \frac{\mathit{\alpha}}{2}\mathit{{l}_{P}}{\left({{\vec \nabla}\times{{\vec B}}}\right)}_{k} - \mathit{\beta}{\partial}_{t}{E}_{k}}\right),   \label{G_H20}
\end{equation}

\begin{equation}
{\tilde{{\cal F}}}_{0i} = {B}_{i} - \frac{\mathit{\alpha}}{2}\mathit{{l}_{P}}{\left({{\vec \nabla}\times{{\vec B}}}\right)}_{i} - \mathit{\beta}\mathit{{l}_{P}}{\partial}_{t}{E}_{i},  \label{G_H25}
\end{equation}

\begin{equation}
{\tilde{{\cal F}}}_{ij} = {\varepsilon}_{ijk}\left({{E}_{k} - \left({ - \frac{\mathit{\alpha}}{2} + \mathit{\beta}}\right)\mathit{{l}_{P}}{\left({\vec \nabla}\times{{\vec E}}\right)}_{k}}\right).  \label{G_H30}
\end{equation}

Having characterized the electromagnetic field-strength tensor, we can now consider a new effective Lagrangian density.

\subsection{Effective Lagrangian and Gambini-Pullin equations}

In this case, the corresponding model is governed by the Lagrangian density:
\begin{equation}
{\cal L} = - \frac{1}{4}{\cal F}_{\mathit{\mu}\mathit{\nu}}^{2} + \frac{\mathit{\xi}}{2} \ \mathit{{l}_{P}}{n}_{\mathit{\kappa}}{\varepsilon}^{\mathit{\kappa}}{}_{\mathit{\alpha}\mathit{\beta}\mathit{\gamma}}\ {n}^{\mathit{\lambda}}{\tilde{F}}_{\mathit{\lambda}}{}^{\mathit{\alpha}}  {\partial}^{\mathit{\beta}}{n}^{\mathit{\rho}}{\tilde{F}}_{\mathit{\rho}}{}^{\mathit{\gamma}} + {a} \ \mathit{{l}_{P}}{n}^{\mathit{\kappa}}{n}^{\mathit{\lambda}}{n}^{\mathit{\rho}}{\partial}_{\mathit{\kappa}}{\tilde{F}}_{\mathit{\lambda}\mathit{\sigma}}{\partial}_{\mathit{\rho}}{A}^{\mathit{\sigma}}, \label{G_H35}
\end{equation}
where $\xi$ and $a$ are dimensionless numerical coefficients.

On simplification (after expansion in ${l}_{P}$) we find that equation (\ref{G_H35}) can be brought into the form
\begin{eqnarray}
{\cal L}&=& - \frac{1}{4}{F}_{\mathit{\mu}\mathit{\nu}}^{2} - \frac{\mathit{\alpha}}{2}\mathit{{l}_{P}}\,{F}^{\mathit{\mu}\mathit{\nu}}{n}_{\mathit{\kappa}}{\varepsilon}^{\mathit{\kappa}}{}_{\mathit{\mu}\mathit{\alpha}\mathit{\beta}}{\partial}^{\mathit{\alpha}}{F}_{\mathit{\nu}}{}^{\mathit{\beta}} +\frac{\mathit{\xi}}{2} \ \mathit{{l}_{P}}\,{n}_{\mathit{\kappa}}{\varepsilon}^{\mathit{\kappa}}{}_{\mathit{\alpha}\mathit{\beta}\mathit{\gamma}} \ {n}^{\mathit{\lambda}}{\tilde{F}}_{\mathit{\lambda}}{}^{\mathit{\alpha}}{\partial}^{\mathit{\beta}}{n}^{\mathit{\rho}}{\tilde{F}}_{\mathit{\rho}}{}^{\mathit{\gamma}} \nonumber\\
&+& a\, \mathit{{l}_{P}}{n}^{\mathit{\kappa}}{n}^{\mathit{\lambda}}{n}^{\mathit{\rho}}{\partial}_{\mathit{\kappa}}{\tilde{F}}_{\mathit{\lambda}\mathit{\sigma}}{\partial}_{\mathit{\rho}}{A^\sigma }. \label{G_H40}
\end{eqnarray}
Notice that to get this new equation, we have ignored quadratic terms in ${l}_{P}$, as it shall become clear below.
In other words, this new effective theory provides us with a suitable starting point for our analysis.

The equations of motion following from the Lagrangian density (\ref {G_H40}) read
\begin{equation}
{\partial}_{\mathit{\mu}}{F}^{\mathit{\mu}\mathit{\nu}} - \mathit{{l}_{P}}\left[{\left({\mathit{\alpha} - \mathit{\xi}{n}^{2}}\right)\Delta + \left({\mathit{\xi} + {2}{a}}\right){\left({{n}\cdot\partial}\right)}^{2}}\right]{n}_{\mathit{\kappa}}{\tilde{F}}^{\mathit{\kappa}\mathit{\nu}} = 0, \label{G_H45}
\end{equation}
where $\Delta  \equiv {\partial _\mu }{\partial ^\mu }$.

From equation (\ref {G_H45}), it now follows, for a time-like four-vector ${n}_{\mathit{\mu}}{=}\left({1,{\vec 0}}\right)$
and $\nu=0$, that
\begin{equation}
\nabla\cdot{\vec E} = 0, \label{G_H50}
\end{equation}
while for $\nu=i$, we find
\begin{equation}
(\nabla\times{\vec B}) + \mathit{{l}_{P}}(\nabla\times\nabla\times{\vec B}) = \frac{\partial}{\partial{t}}{\vec E}.  \label{G_H55}
\end{equation}
It should be further noticed that, to obtain the foregoing equations, we have adjusted the dimensionless numbers ($\alpha$, $a$ and $\xi$) so that they satisfy the algebraic equations $\mathit{\alpha} + {2}{a} = 0$ and $\mathit{\xi} - \mathit{\alpha} = -1$.

Before going on, a remark is pertinent at this point. Observe that equation (\ref{G_H10}) yields
\begin{eqnarray}
{\partial}_{\mathit{\mu}}{\tilde{\cal F}}^{\mathit{\mu}\mathit{\nu}}&=&{\partial}_{\mathit{\mu}}{\tilde{F}}^{\mathit{\mu}\mathit{\nu}} - \mathit{{l}_{P}}\left({ -\frac{\mathit{\alpha}}{2} + \mathit{\beta}}\right){n}_{\mathit{\kappa}}\Delta{F}^{\mathit{\kappa}\mathit{\nu}} + \mathit{{l}_{P}}\left({ - \frac{\mathit{\alpha}}{2} + \mathit{\beta}}\right){n}_{\mathit{\kappa}}{\partial}^{\mathit{\nu}}{\partial}_{\mathit{\mu}}{F}^{\mathit{\kappa}\mathit{\mu}} \nonumber\\
&+&\mathit{{l}_{P}}\frac{\mathit{\alpha}}{2}{n}^{\mathit{\mu}}{\partial}^{\mathit{\kappa}}{\partial}_{\mathit{\mu}}{F}_{\mathit{\kappa}}{}^{\mathit{\nu}}.  \label{G_H60}
\end{eqnarray}
Here, we clearly see that if the right-hand side of equation (\ref{G_H60}) is zero, we re-obtain the Bianchi identities for the new field-strength (${\tilde{\cal F}}^{\mathit{\mu}\mathit{\nu}}$). To examine this question let us recall that 
${\partial}_{\mathit{\mu}}{\tilde{F}}^{\mathit{\mu}\mathit{\nu}} = {0}$. Second, it should be noticed that, for theories described by ${\cal L} = {-}\frac{1}{4}{\cal F}_{\mathit{\mu}\mathit{\nu}}{\cal F}^{\mathit{\mu}\mathit{\nu}}$, the equations of motion read ${\partial}_{\mathit{\mu}}{F}^{\mathit{\mu}\mathit{\nu}}\sim{\cal O}\left({\mathit{{l}_{P}}}\right)$. With this information, it is evident that the third and fourth terms on the right-hand side of equation (\ref{G_H60}) are terms ${\cal O}\left({{\left({{l}_{P}}\right)}^{2}}\right)$. It is also of interest to remark that, from the Bianchi identity, $
{\partial}_{\mathit{\mu}}{F}_{\mathit{\nu}\mathit{\kappa}} + {\partial}_{\mathit{\nu}}{F}_{\mathit{\kappa}\mathit{\mu}} + {\partial}_{\mathit{\kappa}}{F}_{\mathit{\mu}\mathit{\nu}} = {0}
$, it follows that $\Delta{F}_{\mathit{\nu}\mathit{\kappa}}\sim{\cal O}\left({\mathit{{l}_{P}}}\right)$. Making use of the foregoing results, we can, therefore, write (to order ${l}_{P}$)
\begin{equation}
{\partial}_{\mathit{\mu}}{\tilde{\cal F}}^{\mathit{\mu}\mathit{\nu}} = 0.  \label{G_H65}
\end{equation}

Having made these observations, we can immediately write the field equations for a background time-like four-vector 
${n}_{\mathit{\mu}}{=}\left({1,{\vec 0}}\right)$. When $\nu=0$, we obtain
\begin{equation}
\nabla\cdot{\vec B} = 0,  \label{G_H70}
\end{equation}
whereas for $\nu=i$,
\begin{equation}
(\nabla\times{\vec E}) + \mathit{{l}_{P}}\,(\nabla\times\nabla\times{\vec E}) = - \frac{\partial}{\partial{t}}{\vec B}, \label{G_H75}
\end{equation}
to get this last line, we have used the algebraic equation 
$- \frac{\mathit{\alpha}}{2} + \mathit{\beta} = - 1$.\\

In summary, then: with the aid of the equations (\ref {G_H55}) and (\ref {G_H75}), we obtain the equations
\begin{equation}
\frac{\partial}{\partial{t}}{\vec E} = \left({\nabla\times{\vec B}}\right) - \mathit{{l}_{P}}\,{\nabla}^{2}{\vec B}, \label{G_H80}
\end{equation}
and
\begin{equation}
\frac{\partial}{\partial{t}}{\vec B} = - \left({\nabla\times{\vec E}}\right) + \mathit{{l}_{P}}\,{\nabla}^{2}{\vec E}.\label{G_H85} 
\end{equation}
It should again be stressed here that, as discussed previously, the preceding equations are valid at leading order in ${l}_{P}$. This result agrees with that of Ref. \cite{Gambini:1998it,Alfaro:2001rb}, and finds here an independent derivation.

\subsection{Properties}

\subsubsection{Birefringence}

For the sake of completeness, we now re-examine the propagation of electromagnetic waves in this scenario. As well known, in this theory the phenomenon of birefringence is present. We recall, in passing, that birefringence refers to the property that polarized light in a particular direction travels at a different velocity from that of light polarized in a direction perpendicular to this axis. Incidentally, it is of interest to notice that due to quantum fluctuations. the vacuum of this new effective theory has this property, as we are going to show.

To illustrate this important feature, we introduce a decomposition into plane waves for the fields $\vec E$ and $\vec B$ as: 
\begin{equation}
{\vec E} = {\vec E}_{0}{e}^{{-}{i}\left({{wt}{-}{\vec k}\cdot{\vec x}}\right)} , \ {\vec B} = {\vec B}_{0}{e}^{{-}{i}\left({{wt}{-}{\vec k}\cdot{\vec x}}\right)}.  \label{G_H86} 
\end{equation}
These equations can alternatively be rewritten in the form
\begin{equation}
{\vec E} = {\vec E}_{0R}\cos\left({{\vec k}\cdot{\vec x} - {wt}}\right) - {\vec E}_{0I}\sin\left({{\vec k}\cdot{\vec x} - {wt}}\right), \label{G_H90a} 
\end{equation}
and
\begin{equation}
{\vec B} = {\vec B}_{0R}\cos\left({{\vec k}\cdot{\vec x} - {wt}}\right) - {\vec B}_{0I}\sin\left({{\vec k}\cdot{\vec x} - {wt}}\right). \label{G_H95a} 
\end{equation}

Now, making use of equations  (\ref{G_H80}) and (\ref{G_H85}), we readily find that 
\begin{equation}
{- {\vec k}\times{\vec E}_{0R} - {k}^{2} \ {\vec E}_{0I} = - w \ {\vec B}_{0R}}, \label{G_H100a} 
\end{equation}
\begin{equation}
{- {\vec k}\times{\vec E}_{0I} + {k}^{2} \ {\vec E}_{0R} = - w \ {\vec B}_{0I}}, \label{G_H100b} 
\end{equation}
\begin{equation}
{- {\vec k}\times{\vec B}_{0R} - {k}^{2} \ {\vec B}_{0I} = w \ {\vec E}_{0R}}, \label{G_H100c} 
\end{equation}
\begin{equation}
{-{\vec k}\times{\vec B}_{0I} + {k}^{2} \ {\vec B}_{0R} = w \ {\vec E}_{0I}}. \label{G_H100d} 
\end{equation}

Next, without loss of generality we take the $z$-axis as the direction of propagation of the wave light, that is, ${\vec k} = k\hat z$. We further consider a circularly polarized light wave, ${{\vec E}_{0R}} = {{\vec E}_0}\hat x$ and ${{\vec E}_{0I}} = \eta {{\vec E}_0}\hat y$, where $\eta  =  \pm 1$, is a parameter to describe the fields with the left- and right-hand circularly polarized unit vectors. 

Now, by using (\ref{G_H100a}) - (\ref{G_H100d}), we find that the corresponding dispersion relation becomes 
\begin{equation}
{\omega ^2} = {k^2} + 2{l}_{P} \eta {k^3} + {l}_{P}^{2}{k^4}. \label{G_H105}
\end{equation}
Expression (\ref{G_H105}) immediately shows that the parameters $\eta$ and ${l}_{P}$ are coupled or, more precisely, the term ${l}_{P}^{2}$ does not distinguish between the two modes of polarization. Here is the physical reason why, in this work, we have considered our calculations at leading order in ${l}_{P}$.

It is immediate to check that, for ${l}_{P}^{2} \ll {l}_{P}$, we recover the known result \cite{Gambini:1998it,Alfaro:2001rb}
\begin{equation}
{\omega ^2} = {k^2} + 2{l}_{P} \eta {k^3}. \label{G_H110a}
\end{equation}
As we have already mentioned, analogous dispersion relation can be obtained from a different effective theory with Lorentz-violating dimension-$5$ operators for the photon sector \cite{Myers:2003fd}.
This implies that the electromagnetic waves with different polarizations have different velocities, that is,
the vacuum birefringence phenomenon is present.
Let us also note here that the group velocity reduces to
\begin{equation}
{v_g} = \frac{1}{\omega }\left( {k + 3\eta {l}_{P}\,{k^2}} \right), \label{G_H110b}
\end{equation}
which displays a modified speed of light for a photon with momentum $k$. 

\subsubsection{The energy-momentum tensor}

Another relevant aspect to explore in this new electrodynamics is the calculation of the energy and momentum densities transported by the electromagnetic field.
To do this, we start by contracting the equations of motion (\ref{G_H45}) with $F_{\nu \kappa }$, that is,
\begin{eqnarray}
\left( {{\partial _\mu }{F^{\mu \nu }}} \right){F_{\nu \kappa }} &-& \alpha \,{l}_{P}\,{n_\lambda }{\varepsilon ^{\lambda \nu \alpha \beta }}\left( {{\partial _\alpha }{\partial ^\mu }{F_{\mu \beta }}} \right){F_{\nu \kappa }} + \xi \,{l}_{P}\,{n^2}{n^\rho }\left( {\Delta {{\tilde F}_\rho }^\nu } \right){F_{\nu \kappa }} \nonumber\\
&-& \left( {\xi  + 2a} \right)\,{l}_{P} \left[ {{{\left( {n \cdot \partial } \right)}^2}{n_\rho }{{\tilde F}_\rho }^\nu } \right]{F_{\nu \kappa }} = 0.
 \label{G_H111}
 \end{eqnarray}
After some further algebraic manipulations, for the model under consideration, the energy-momentum tensor comes out in the form
\begin{eqnarray}
{\Theta ^\mu }_\kappa &=& {F^{\mu \nu }}{F_{\nu \kappa }} + \frac{1}{4}{\delta ^\mu }_\kappa {F^2} + {l}_{P}\,{n_\lambda }\left[ {\left( { - \alpha  + \xi {n^2}} \right){\partial ^\mu } - \left( {\xi  + 2a} \right){n^\mu }\left( {n \cdot \partial } \right)} \right]{\tilde F^{\lambda \nu }} {F_{\nu \kappa }} \nonumber\\
&+& \frac{{l}_{P}}{2}\left( { - \alpha  + \xi {n^2}} \right){n_\kappa }\left( {{\partial _\alpha }{A_\nu }} \right)\left( {{\partial ^\alpha }{{\tilde F}^{\mu \nu }}} \right) - \frac{{l}_{P}}{2}\left( {\xi  + 2a} \right){n_\kappa }\left( {n \cdot \partial } \right){A_\nu }\left( {n \cdot \partial } \right){\tilde F^{\mu \nu }}, \nonumber\\
\label{G_H115}
\end{eqnarray}
which is conserved
\begin{equation}
{\partial _\mu }{\Theta ^\mu }_\kappa  = 0. \label{G_H120}
\end{equation}
Note that here, by virtue of the anisotropy described by the external vector, we do not obtain a symmetric energy-momentum tensor
\begin{equation}
{\Theta ^\mu }_\kappa  \ne {\Theta _\kappa }^\mu. \label{G_H125}
\end{equation}

Before we proceed further, two remarks are pertinent at this point, because this energy-momentum tensor is neither symmetric nor manifestly gauge invariant. First, it would be noticed that the last two terms at the right-hand side of expression (\ref{G_H115}) explicitly contain the potential $A_\mu$; hence, expression (\ref{G_H115}) would not be manifestly gauge invariant. In such a case, we verify that, under a gauge transformation, $\mathit{\delta}{A}_{\mathit{\nu}} = {\partial}_{\mathit{\nu}}\Lambda$, the fourth term at the right-hand side of (\ref{G_H115}) transforms as 
 $\frac{\chi }{2}\left( { - \alpha  + \xi {n^2}} \right){n_\kappa }\left( {{\partial _\alpha }{\partial_\nu \Lambda}} \right)\left( {{\partial ^\alpha }{{\tilde F}^{\mu \nu }}} \right)$. After using the Bianchi identity, this term will appear in the continuity equation (\ref{G_H120}) as $
{\partial}_{\mathit{\mu}}{\partial}_{\mathit{\nu}}\left({{\partial}_{\mathit{\alpha}}\Lambda{\partial}^{\mathit{\alpha}}{\tilde{F}}^{\mathit{\mu}\mathit{\nu}}}\right) = 0$. The same argument applies to the fifth term of expression (\ref{G_H115}). Therefore, the continuity equation (\ref{G_H120}) is actually gauge invariant, though not is a manifest way. Second, we wish to further elaborate on the physical meaning of expression  (\ref{G_H125}). Here, our concern is to reconsider the calculation of the energy density and the Poynting vector, paying due attention to the non-symmetric character of the energy-momentum tensor (\ref{G_H125}). For this purpose, we consider the last two terms at the right-hand side of the expression for ${\Theta}^{\mathit{0}\mathit{0}}$, which would spoil a correct definition of energy. As we can see, under a gauge transformation, the fourth term of ${\Theta}^{\mathit{0}\mathit{0}}$ transforms as $\nabla\cdot\left({{\partial}_{\mathit{\alpha}}\Lambda{\partial}^{\mathit{\alpha}}{\vec B}}\right)$.
Thus, when we consider this term in the energy expression ($\int{{d}^{3}x} \ {\Theta}^{00}$), we verify
that it becomes a surface contribution which integrates to zero assuming, as always, that the field decreases sufficiently fast at infinity. The same argument applies to the fifth term of ${\Theta}^{\mathit{0}\mathit{0}}$. Therefore, the energy is gauge invariant. In the same way as done in the previous case, the terms dependent on the gauge parameter that contribute to ${\Theta}^{\mathit{0}\mathit{i}}$ disappear. Thus, the momentum, ${\vec P} = \int{{d}^{3}x}{\Theta}^{0i}$, is also a gauge invariant quantity.

We now consider a background time-like ${n}^{\mathit{\mu}}{=}\left({1,\vec 0}\right)$. Making use of the foregoing results, we find that
\begin{equation}
{\Theta ^{00}} \equiv {\cal E} = \frac{1}{2}\left( {{{\vec E}^2} + {{\vec B}^2}} \right) - \left( {\alpha  + 2a} \right) {l}_{P} \left( {{\vec E} + \frac{1}{2}\dot {\vec A}} \right) \cdot \dot {\vec B} + \frac{1}{2}\left( {\alpha  - \xi } \right) {l}_{P} \left( {{\partial _j}{A_i}} \right)\left( {{\partial _j}{B_i}} \right),  \label{G_H130}
\end{equation}
where ${\cal E}$ denotes an energy density. While the momentum density, ${\vec {\tilde S}}$, is given by
\begin{equation}
{\Theta ^{0i}} \equiv {\vec {\tilde S}} = {\vec E} \times {\vec B} + \left( {\alpha  + 2a} \right) {l}_{P}\,{\vec B} \times \dot {\vec B}. \label{G_H135}
\end{equation}
We also recall that $\alpha  + 2a = 0$ and $\xi  =  - 1 + \alpha$. Thus, finally we end up with
\begin{equation}
{\cal E} = \frac{1}{2}\left( {{{\vec E}^2} + {{\vec B}^2}} \right) + \frac{{l}_{P}}{2}\left( {{\nabla ^2}{\vec A}} \right) \cdot {\vec B},  \label{G_H140}
\end{equation}
and
\begin{equation}
{\vec {\tilde S}} = {\vec S}_{Maxwell} = {\vec E} \times {\vec B}, \label{G_H145}
\end{equation}
which reduces to the usual Poynting vector.

\section{Extension to a space-like external LSV four-vector}

Proceeding in the same way as we did in the previous Section, we shall now consider the field equations for a background space-like four-vector $n^{\mu}$, that is, ${n}^{\mathit{\mu}} = \left({0,\vec n}\right)$.

\subsection{Maxwell-type field equations}
In other words, we now wish to obtain equations (\ref{G_H80}) and (\ref{G_H85}) for a background space-like four-vector. 

For this, we restrict our attention to the equation (\ref{G_H45}). We then obtain for $\nu=0$
\begin{equation}
{\partial}_{i}{F}^{i0} - \mathit{{l}_{P}}\left[{\left({\mathit{\alpha} + \mathit{\xi}}\right)\Delta + \left({\mathit{\xi}+ {2}{a}}\right){\left({{\vec n}\cdot\nabla}\right)}^{2}}\right]{n}_{j}{\tilde{F}}^{j0} = 0. \label{noni05}
\end{equation}
While for $\nu=i$
\begin{equation}
{\partial}_{0}{F}^{0i} + {\partial}_{j}{F}^{ji} - \mathit{{l}_{P}}\left[{\left({\mathit{\alpha} + \mathit{\xi}}\right)\Delta{+}\left({\mathit{\xi} + {2}{a}}\right){\left({{\vec n}\cdot\nabla}\right)}^{2}}\right]{n}_{j}{\tilde{F}}^{ji} = {0}.  \label{noni10}
\end{equation}
It is of interest also to notice that, for $\nu=0$, we arrive at
\begin{equation}
{\tilde{\cal F}}_{0i} = {\tilde{F}}_{0i} + \left({\frac{\mathit{\alpha}}{2} - \mathit{\beta}}\right)\mathit{{l}_{P}}\,{n}^{k}{\partial}_{t}{F}_{ki} - \left({\frac{\mathit{\alpha}}{2} - \mathit{\beta}}\right)\mathit{{l}_{P}}\,{n}^{k}{\partial}_{i}{F}_{k0}, \label{noni15}
\end{equation}
whereas for $\nu=i$
\begin{equation}
{\tilde{\cal F}}_{ij} = {\tilde{F}}_{ij} + \left({\frac{\mathit{\alpha}}{2} - \mathit{\beta}}\right)\mathit{ {l}_{P}}\,{n}^{k}{\partial}_{i}{F}_{kj} - \left({\frac{\mathit{\alpha}}{2} - \mathit{\beta}}\right)\mathit{{l}_{P}}\,{n}^{k}{\partial}_{j}{F}_{ki}.\label{noni15}
\end{equation}

Based on these equations, we can finally present the following Maxwell-type equations:
\begin{equation}
\nabla\cdot{\vec E} + \mathit{{l}_{P}} \left[{\left({\mathit{\alpha} + \mathit{\xi}}\right)\Delta + \left({\mathit{\xi} + {2}{a}}\right){\left({{\vec n}\cdot\nabla}\right)}^{2}}\right]\left({{\vec n}\cdot{\vec B}}\right) = 0, \label{noni20}
\end{equation}
\begin{eqnarray}
\nabla\times{\vec E} + \left({\frac{\mathit{\alpha}}{2} - \mathit{\beta}}\right)\mathit{{l}_{P}}\nabla\left({\vec n{\cdot\partial}_{t}\vec E}\right) &=& - {\partial}_{t}{\vec B} - \left({\frac{\mathit{\alpha}}{2} - \mathit{\beta}}\right)\mathit{{l}_{P}} \ {\vec n}{\times\partial}_{t}^{2}{\vec B} \nonumber\\
&+& \left({\frac{\mathit{\alpha}}{2} - \mathit{\beta}}\right)\mathit{{l}_{P}} \ {\vec n} \ {\nabla}^{2}\left({{\vec n}\times{\vec B}}\right) \nonumber\\
&+& \left({\frac{\mathit{\alpha}}{2} - \mathit{\beta}}\right)\mathit{{l}_{P}}\nabla \ \nabla\cdot\left({{\vec n}\times{\vec B}}\right),\label{noni25}
\end{eqnarray}
\begin{equation}
\nabla\cdot{\vec B} + \left({\frac{\mathit{\alpha}}{2} - \mathit{\beta}}\right)\mathit{{l}_{P}} \ \nabla\cdot\left({\vec n{\times\partial}_{t}{\vec B}}\right) + \left({\frac{\mathit{\alpha}}{2} - \mathit{\beta}}\right)\mathit{{l}_{P}}{\nabla}^{2}\left({{\vec n}\cdot{\vec E}}\right) = 0, \label{noni30}
\end{equation}
\begin{equation}
\nabla\times{\vec B} - \mathit{{l}_{P}}\left[{\left({\mathit{\alpha} + \mathit{\xi}}\right)\Delta + \left({\mathit{\xi} + {2}{a}}\right){\left({{\vec n}\cdot\nabla}\right)}^{2}}\right]\left({{\vec n}\times{\vec E}}\right) = {\partial}_{t}{\vec E}.\label{noni35}
\end{equation}
Evidently, these field equations are more involved than those of the (time-like) previous Section. However, we do not intend to pursue a study of these equations in the present contribution. A fuller account on these issues shall be presented elsewhere.

\section{Concluding Remarks}

In summary, we have proposed and studied the properties of a new non-Maxwellian electrodynamics coupled to a Lorentz-violating background through the presence of higher-derivative terms. This effective theory has allowed us to recover the known results obtained in \cite{Gambini:1998it,Alfaro:2001rb} within the Minkowski space covariant formulation of a classical field theory. As already expressed, the benefit of considering the present Lorentz-symmetry violating electrodynamical action is to provide connections among different effective models. Starting off from the situation of a time-like background vector, we made sure that, with our covariant formulation, we could reproduce the Gambini-Pullin's proposal of an electrodynamics accounting for spin-foam effects. That was a sort of validation test of our covariant formulation. Once that point was set up, we have considered the possibility to extend this covariant formulation to include the case Lorentz-symmetry breaking is realized by a space-like vector. This yields to a broad family of models, which encompasses situations in which even the homogeneous Maxwell equations are modified. In the Gambini-Pullin model, there is actually an extra term in the Faraday-Lenz magnetic induction equation; the Gauss equation for the magnetic field remains however the same as in the Maxwellian case. In our space-like extended model, since we have a number of parameters at our disposal, we may end up with both Faraday-Lenz and magnetic-Gauss equations both modified by the Planck scale effect, as shown in the set of equations (\ref{noni20}) - (\ref{noni35}). It is not our aim, in this contribution, to pick out specific choices of parameters, yielding particular actions, to discuss the physical aspects and the consistency (absence of supraluminal signals and non-existence of ghost-type excitations in the second-quantized version) of the models that differ from one another by different choices of parameters in eqs. (\ref{noni20}) - (\ref{noni35}). It is a relevant issue - and special attention shall be devoted - to study the wave equations for the electric and magnetic fields in the case the would-be homogeneous Maxwell equations, (\ref{noni25}) and (\ref{noni30}), exhibit the extra terms with the external space-like vector. Another issue we are going to exploit is the interference between non-linearity and Planck scale effects. If we consider high intensity electric and magnetic fields, such as those produced in extreme light experiments, and Planck scale effects are accounted for, we can inspect how the latter may affect phenomena like photon splitting, light-light Delbr\"uck scattering and the Breit-Wheeler effect. These are topics in our agenda of forthcoming works, which we shall be reporting elsewhere.

{\bf Acknowledgments}: 
One of us (P. G.) was partially supported by Fondecyt (Chile) grant 1180178 and by Proyecto Basal FB0821.

\end{document}